\documentclass[twocolumn,floats,aps]{revtex4}
\usepackage{times,amsmath,amsfonts,mathrsfs,psfrag,latexsym,pstricks,graphics}

\begin{document}
\title{Griffiths-McCoy Singularities in the Dilute Transverse-Field Ising Model:  A Numerical  Linked Cluster Expansion Study}
\author{Foster Thompson}
\affiliation{Case Western Reserve University, OH 44106, USA}
\author{Rajiv R. P. Singh}
\affiliation{University of California Davis, CA 95616, USA}

\date{August 2018}
\begin{abstract}
We use Numerical Linked Cluster Expansions (NLCE) to study the site diluted transverse-field Ising model
on the square-lattice at $T=0$.  NLCE with a self-consistent mean-field on the boundary of the clusters is used to obtain the ground state magnetization, susceptibility, and structure factor as a function of transverse field $h$ and exchange constant $J$.  Adding site-dilution to the model turns NLCE into a series expansion in the dilution parameter $p$.  Studying the divergence of the structure factor allows us to establish the phase-diagram in the $h/J$ and $p$ plane.  By studying the magnetization of the system in a longitudinal field, we investigate the Griffiths-McCoy (GM) singularities.  We find that the magnetization develops non-linearities in the Griffiths phase with exponents that vary continuously with $h$.  
Additionally, the probability distribution of the local susceptibility develops long tails in the Griffith's phase, which is studied in terms of its moments.
\end{abstract}
\maketitle

\section{Introduction}
Disorder occurs in physical systems for a plethora of reasons.  We considered the case of site dilution--the random omission of lattice sites--in the transverse field Ising model in two dimensions.  This models magnetic substances with a fixed concentration of non-magnetic impurities and alloys of magnetic and non-magnetic substances.  Among other properties, this model displays two noteworthy features: a discontinuity in the critical value of the transverse field as a function of the dilution parameter and Griffths-McCoy singularities.

The discontinuity in the transverse field as the dilution parameter varies below the percolation threshold was first
shown by Harris \cite{Harris}.  Below the percolation threshold the lattice is composed of disconnected pieces, so no overall long range order is possible.  Once the lattice percolates, a cluster of spins spanning the entire length of the lattice arises, permitting long range order up to a critical field of at least that of the one-dimensional pure system.  This picture was later confirmed by 
Stinchcombe using a real-space Renormalization Group calculation \cite{Stinchcombe}.

Griffiths Singularities arise due to the low but nonzero probability of large non-dilute regions in an otherwise dilute lattice.  These regions tend to magnetize, locally entering a ferromagnetic phase despite the disordered 
behavior of the bulk lattice, effectively behaving as embedded, finite-size copies of the pure system \cite{Griffiths}.  
These clusters lead to ``Griffiths" singularities everywhere where the pure system is ordered.  
In classical systems, these singularities are weak as the probability of having
a large ordered region falls off exponentially with size
and consequently they are barely visible experimentally or numerically.  
However, in dynamical properties of quantum systems, tunneling between the ground state and excited states in the 
locally non-dilute regions gives the weight of singularities an additional factor related to the inverse of energy gap between these states, which is also exponentially small in the region size. This can promote what were weak
essential singularities in thermodynamic quantities into power-laws.  As a consequence, these quantum Griffiths-McCoy 
singularities \cite{McCoy1,McCoy2} are important both for experiments \cite{experiments} and numerical calculations \cite{rieger,guo,ikegami}.  For a more comprehensive discussion on Griffiths-McCoy singularities in a myriad of disordered models and on the role of disorder in general in quantum critical behavior, see \cite{Vojta}.

In this work, we study the critical behavior and Griffiths-McCoy singularities in the dilute transverse-field Ising model \cite{senthil,ikegami} at $T=0$ using numerical linked cluster expansions (NLCE) \cite{nlc1,nlc2,nlc3,nlc4}.  We use NLCE to study the susceptibility, magnetization, and structure factor and establish the
phase diagram of the system.  We pay special attention to the existence of Griffiths-McCoy singularities, finding numerical evidence for their existence in the behavior of the magnetization as a function of longitudinal field and the probability distribution of the local susceptibility.

\section{Overview of Approach}
\subsection{Model}
We consider the zero temperature behavior of a site-diluted transverse field Ising model on a two-dimensional square lattice.  The model is parametrized by three values $\{J,h,p\}$, where $J$ is an exchange constant that controls the strength of nearest-neighbor spin-spin interactions, $h$ is the coupling strength to a transverse field, and $p$ is the dilution parameter.
The Hamiltonian of the model is:
\begin{equation}
\mathscr{H}=J\sum\limits_{\langle i,j\rangle}\epsilon_i\epsilon_j\sigma_i^z\sigma_j^z+h\sum\limits_i\epsilon_i\sigma_i^x,
\end{equation}
where $\langle\cdot,\cdot\rangle$ denotes nearest-neighbor pairs of sites on the lattice and $\sigma^z$ and $\sigma^x$ are the Pauli matrices.  The $\epsilon_i$ terms are site dilution variables: quenched random variables with bimodal distribution, taking values of $0$ and $1$ with probability $p$ and $1-p$ respectively.

At zero temperature, thermal averages reduce to ground state expectation values.  We denote the ground state by $|0\rangle$.
We will take $J<0$ to study the ferromagnetic problem.  The ground state only depends on the ratio $h/J$. Hence, we set $J=-1$.

\subsection{Method}
Numerical Linked Cluster Expansions (NLCE) is a method of approximating an extensive property of a thermodynamic system as a series with terms computed by exact diagonalization of small, finite size systems---`clusters'---which embed into the lattice of the full model.
%Cite something
Specifically, given an extensive property $P$, its per-site value in the thermodynamic limit is given by the expression:
\begin{equation}
\lim\limits_{N\to\infty}\frac{P}{N}=\sum\limits_c L[c]W[c].
\end{equation}
The lattice constant $L[c]$ denotes the number of ways the cluster $c$ can embed into the lattice.  $W[c]$ is the weight of the cluster in the lattice, determined recursively by:
\begin{equation}
W[c]=P[c]-\sum\limits_{c'\subset c}W[c'],
\end{equation}
where, $c'\subset c$ is a sub-cluster---a cluster which embeds into the cluster $c$---and $P[c]$ is the property computed on the cluster.

For disordered models, one is typically interested in quantities of the form $[P/N]_\text{av}$, where $[...]_\text{av}$ denotes the quenched average. In our case, this is a sum over the site dilution variables $\epsilon_i$.  NLCE can be generalized to a quenched average in a straightforward way by simply computing the quenched average of each cluster individually before summing up the total value of the property \cite{nlcq1,nlcq2,nlcq3}.  In the case of the site diluted model we study, this greatly simplifies the resulting series.  For any cluster, any configuration of the site dilution variables $\epsilon_i$ in which any site is omitted will reduce the cluster to a collection of its sub-clusters, resulting in zero weight after the weights of sub-clusters are subtracted away.  Only the single configuration with no dilution survives the sub-graph subtraction.  Consequently, the NLCE of the quenched average of a property $P$ reduces to simply the NLCE of $P$ for the pure system with an additional factor of $p^N[c]$, the probability of the non-dilute configuration:
\begin{equation}
\lim\limits_{N\to\infty}\bigg[\frac{P}{N}\bigg]_\text{av}=\sum\limits_c L[c]W[c]p^{N[c]}\equiv\sum\limits_{n=1}^\infty a_np^n,
\end{equation}
where $N[c]$ denotes the number of sites in $c$.  As a consequence, the NLCE becomes a power series in the dilution parameter $p$.

\section{Pure System Analysis}
We first considered the non-dilute Ising problem (the $p=1$ limit).  We calculate the susceptibility and structure factor, as both of these quantities are known to diverge strongly near the pure system's critical point of $h_c\simeq3.044$.
%Correct?
%Cite something for value?
The structure factor $S$ is defined by:
\begin{equation}
S=\sum\limits_{i,j}\langle\sigma_i^z\sigma_j^z\rangle-\langle\sigma_i^z\rangle\langle\sigma_j^z\rangle,
\end{equation}
where we use the notation $\langle\sigma_i^z\rangle=\langle0|\sigma_z^i|0\rangle$.
In order to compute the susceptibility, we add an additional longitudinal field term to the Hamiltonian:
\begin{equation}
\mathscr{H}=J\sum\limits_{\langle i,j\rangle}\sigma_i^z\sigma_j^z+h\sum\limits_i\sigma_i^x+h_L\sum\limits_i\sigma_i^z.
\end{equation}
With this additional factor, letting $E_0$ denote the ground state energy, the susceptibility is given by:
\begin{equation}
\chi=-\lim\limits_{h_L\to0}\frac{\partial^2E_0}{{\partial h_L}^2}.
\end{equation}

\begin{figure}[htb]
\begin{center}
\scalebox{.335}{\includegraphics{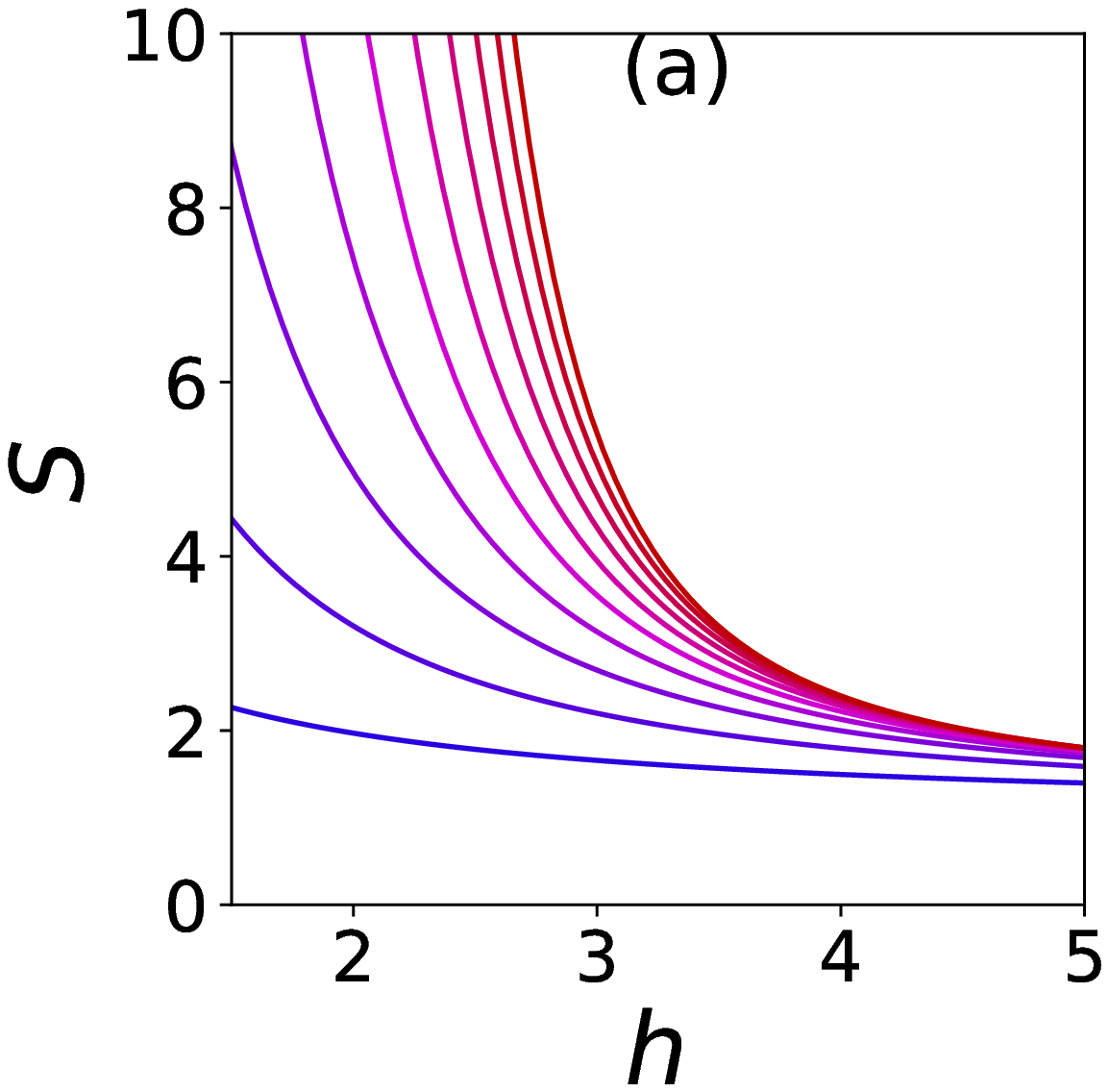}}
\scalebox{.335}{\includegraphics{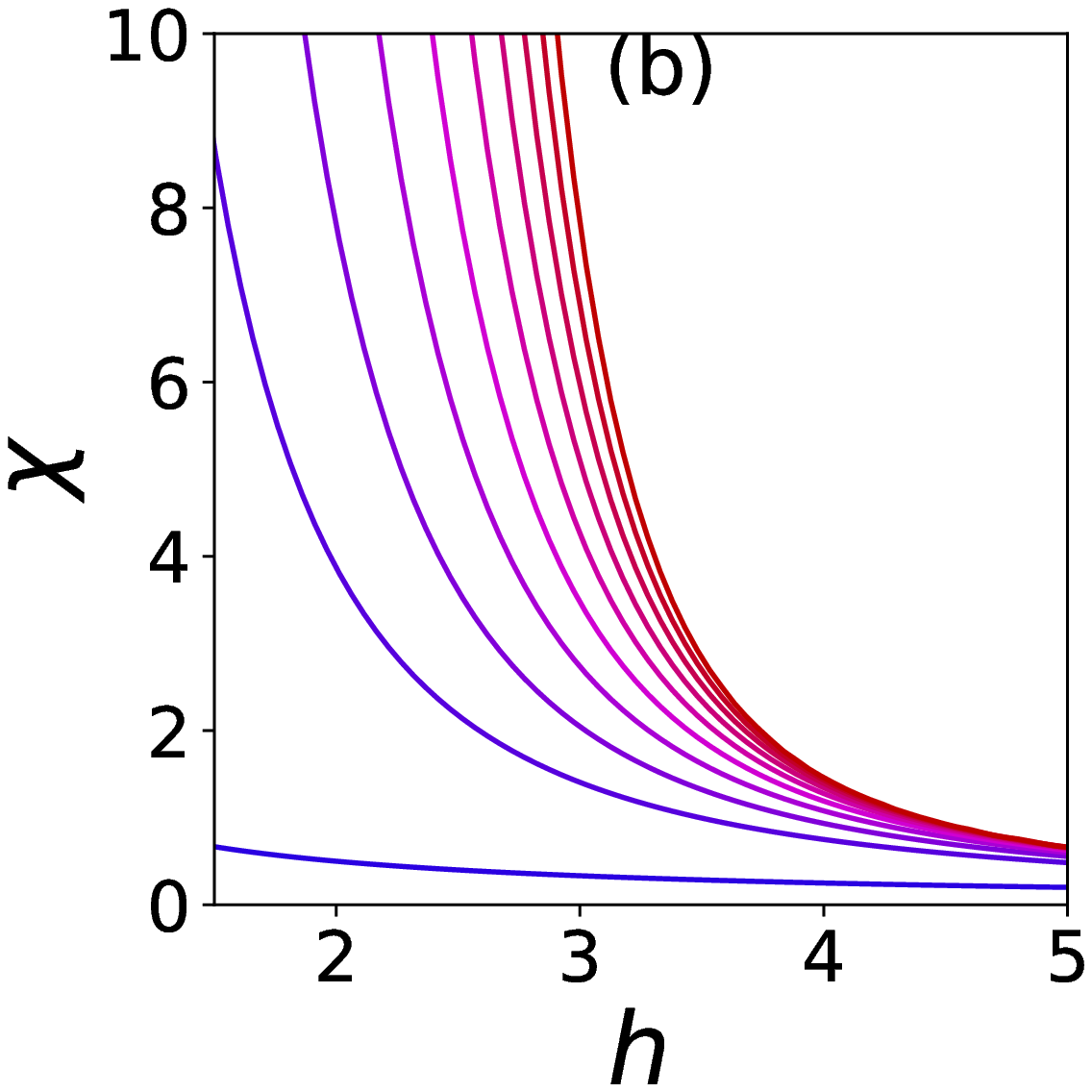}}
\caption{(a) shows the structure factor $S$ and (b), the susceptibility $\chi$.  For both quantities, the NLCE is truncated to include only clusters with 10 or fewer bonds.  Curves of order 1-10 in the number of bonds are shown simultaneously with increasing redness starting with blue at first order.  The NLCE converge well for $h>h_c$ and increase sharply close to $h_c$ before saturating for small values of $h$.  The point at which the steep rise begins occurs closer and closer to $h_c$ the higher the order.}
\label{f:PureSystem}
\end{center}
\end{figure}

\begin{figure}[htb]
\begin{center}
\scalebox{.335}{\includegraphics{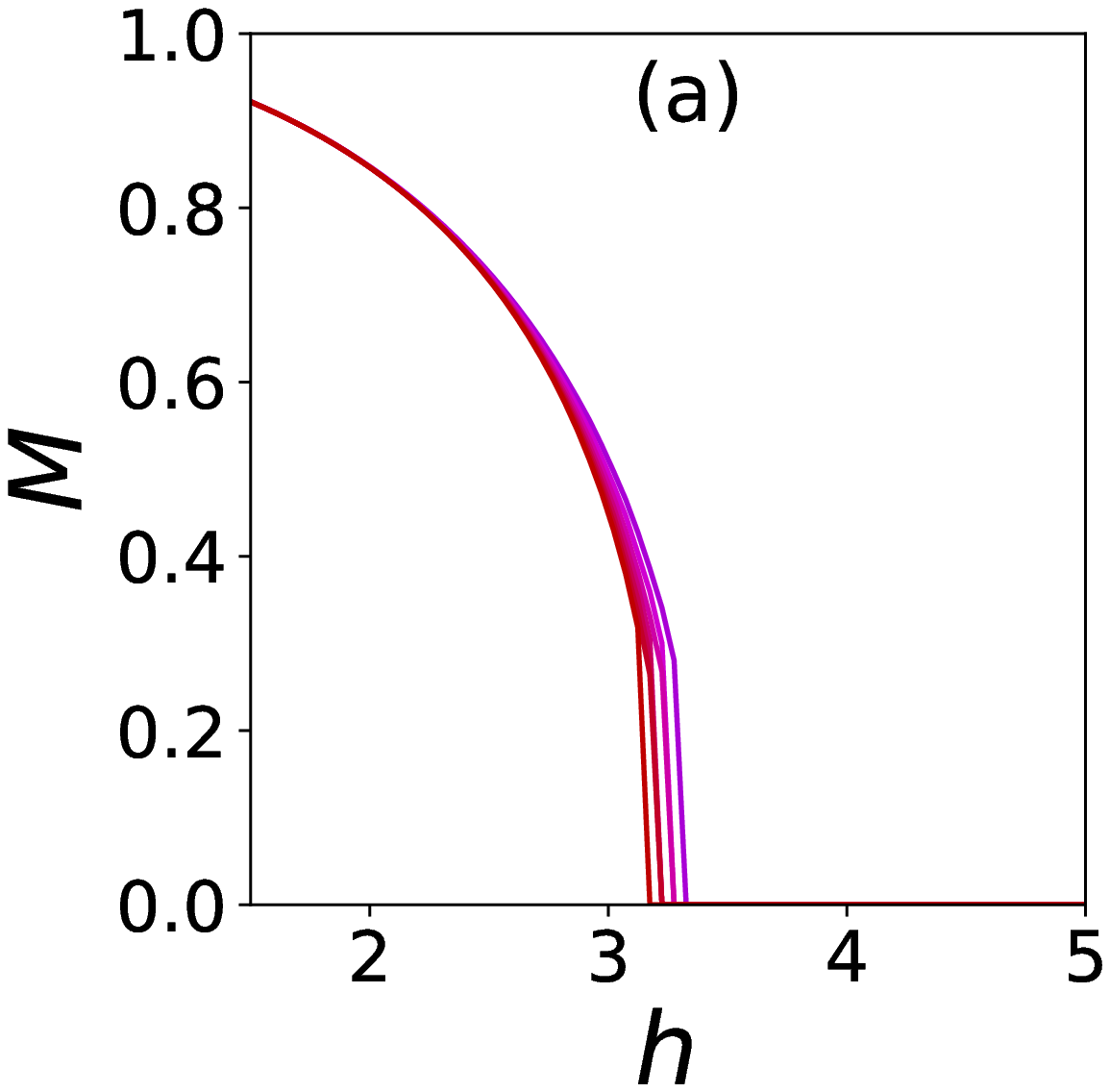}}
\scalebox{.335}{\includegraphics{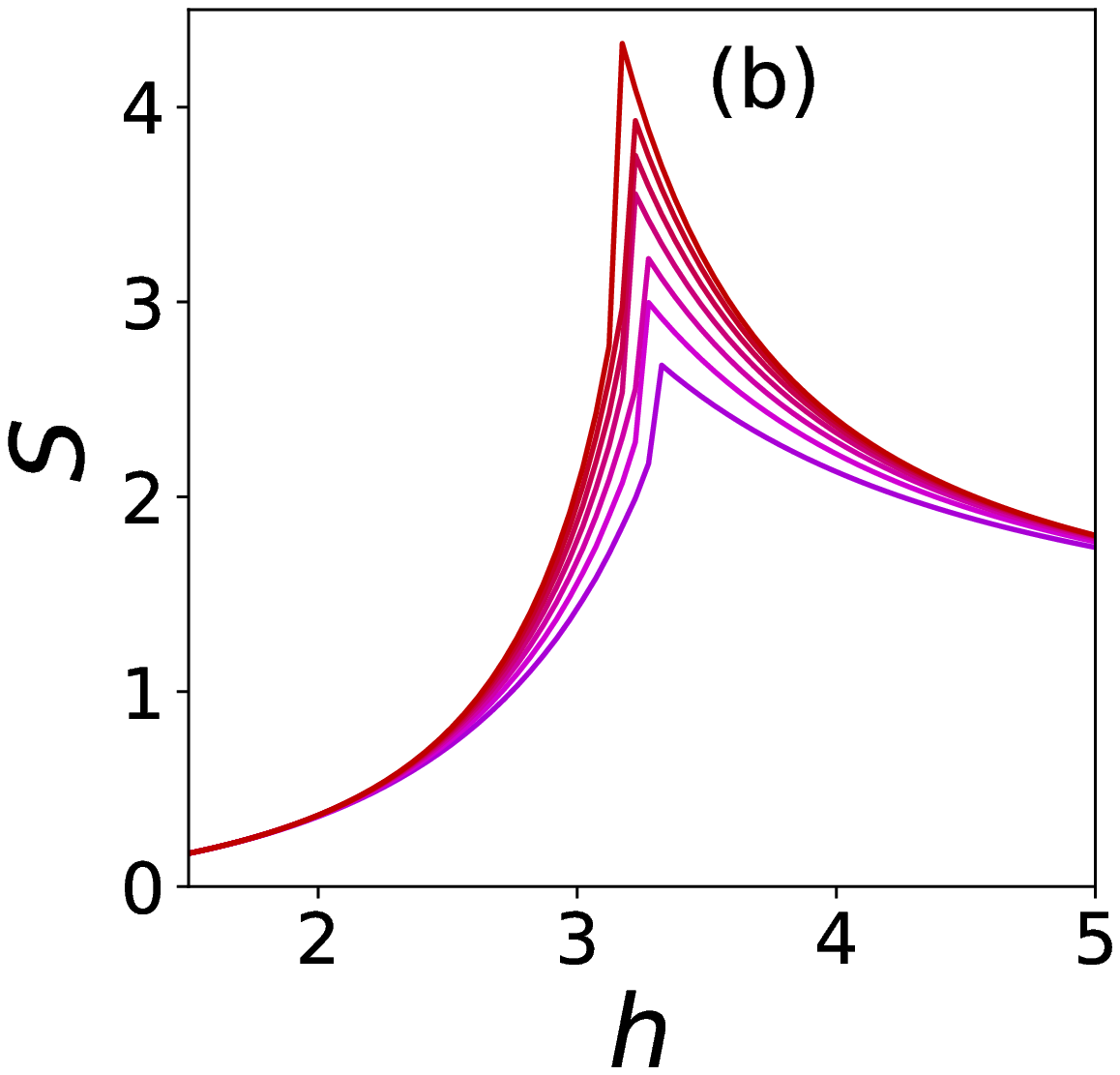}}
\scalebox{.335}{\includegraphics{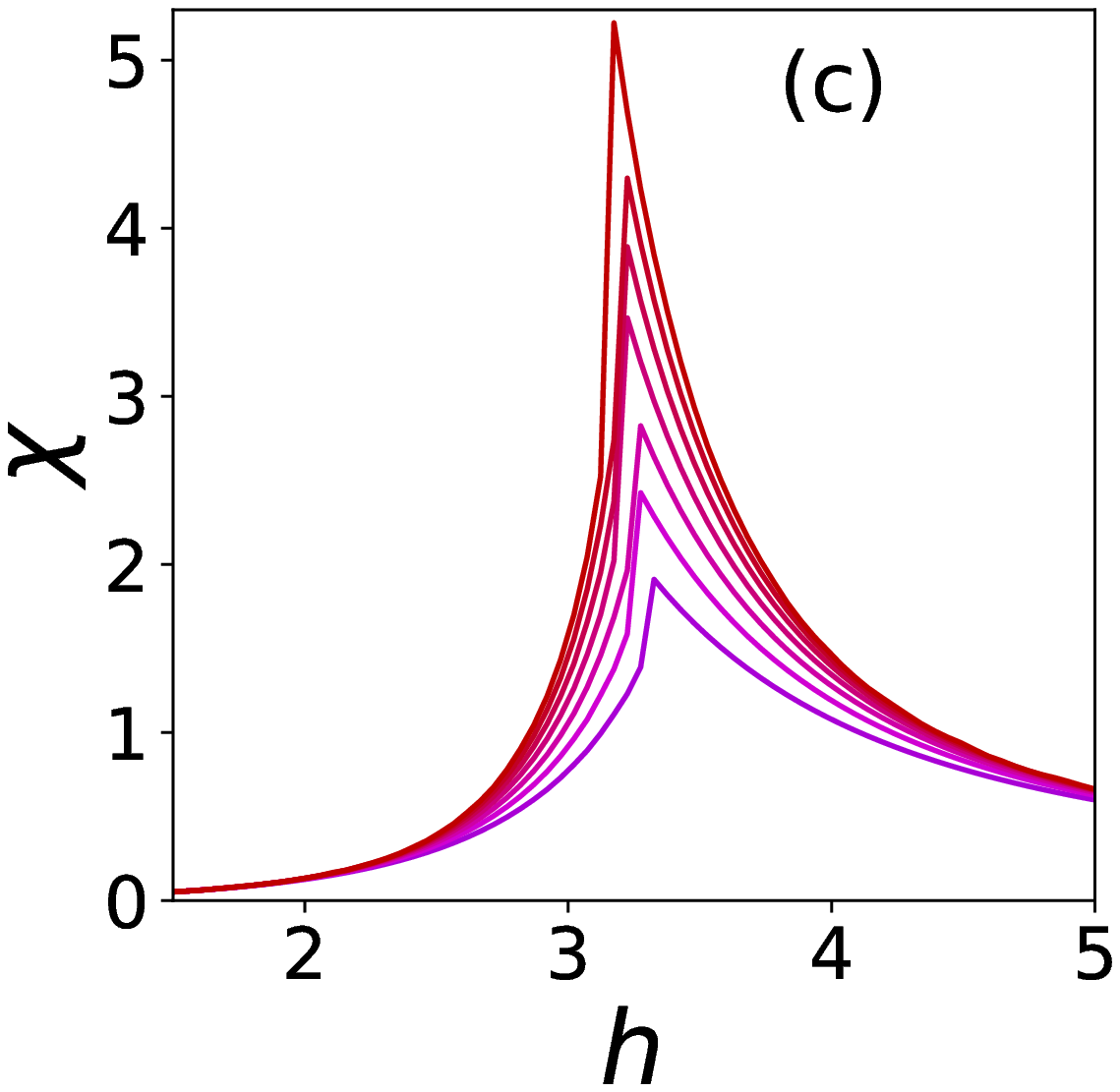}}
\caption{In all three plots, orders 4-10 in the number of bonds are shown, with higher orders shown in redder colors.  (a) shows the magnetization computed using the self-consistency constraint.  Its shape is qualitatively typical for the magnetization of a ferromagnet, falling to zero near the critical point.  (b) and (c) show curves of the structure factor $S$ and susceptibility $\chi$  plotted using the magnetization self-consistent mean field.  Both quantities now converge very well in both the ordered and disordered phases, peaking near the critical point.}
\label{f:PureSystemMF}
\end{center}
\end{figure}

We use NLCE \cite{nlc1,nlc2,nlc3,nlc4} to compute the susceptibility
$\chi$ and structure factor $S$ for a range of values of $h$ on both sides of $h_c$.  The results of the computation are shown in fig. \ref{f:PureSystem}.  
Without help of an extrapolation method it is difficult to locate the critical point in NLCE as the structure factor and susceptibility
keep on increasing as $h$ is reduced in each order.
However, we can do better by including an additional `mean-field' term in the Hamiltonian at the boundary of each cluster to account for the effects of the rest of the lattice:
\begin{equation}
\mathscr{H}=J\sum\limits_{\langle i,j\rangle}\sigma_i^z\sigma_j^z+h\sum\limits_i\sigma_i^x+m(h)\sum_iq_i[c]\sigma_i^z.
\end{equation}
This added term is a mean field acting on the boundary of the cluster: the quantity $q_i[c]$ represents the number of neighbors on the site $i$ that are not in the cluster $c$ and so are outside the cluster.  We expect the $h$-dependent value of $m$ to satisfy the self-consistency condition $m=M$, where $M$ is the magnetization per site of the lattice, defined by:
\begin{equation}
M={1\over N}\sum\limits_i\langle\sigma_i^z\rangle.
\end{equation}
We implement this as a constraint numerically by computing $M$ for a small number of guessed, constant values of $m$ for each $h$, then interpolating to find the approximate value of $m$ satisfying the self-consistency constraint.
%explain better/more?
Computed for a range of $h$, this gives an approximation for the magnetization $M$, and also regulates the convergence of the NLCE for both $S$ and $\chi$.  The result of this computation is shown in fig. \ref{f:PureSystemMF}.  This gives a qualitative picture of the phase transition but is not accurate enough to calculate critical properties. In this manuscript, we are not interested in a more complex extrapolation of the NLCE series, as our primary interest is in the diluted system, which turns out to be much more straightforward to extrapolate.

%Concluding remarks; say more?

\section{Dilution Problem}
\subsection{Phase Diagram}
We now turn our attention to the dilute Ising problem.  In the limit of $h\to0$, this becomes a percolation problem with two phases controlled by the value of the percolation parameter $p$.  At low values of $p$, highly dilute configurations dominate, the lattice becomes a collection of disconnected clusters, and no long-range order develops.  Above the percolation threshold $p_c\simeq0.59$,
%from where?
the lattice percolates, with an infinite cluster spanning the full length of the lattice. This infinite cluster of spins 
can develop long-range order.
For small, nonzero values of $h$, it has been shown \cite{Harris,Stinchcombe}
%Cite Harris and other stuff(RG analysis confirmation from Stinchcombe? what else?)
that a flat phase boundary with $p_c$ independent of $h$ extends into the $h-p$ plane to some value $h_M$ with a lower bound of $h=1$, the critical point of the one-dimensional model.  Near the critical point $h_c$ of the pure system, the phase boundary is believed to extend smoothly downward into the plane before meeting the flat boundary at the multi-critical point $h_M$.  We used NLCE to confirm this picture of the phase diagram, and get an estimate of $h_M$.

\begin{figure}[htb]
\begin{center}
\scalebox{.55}{\includegraphics{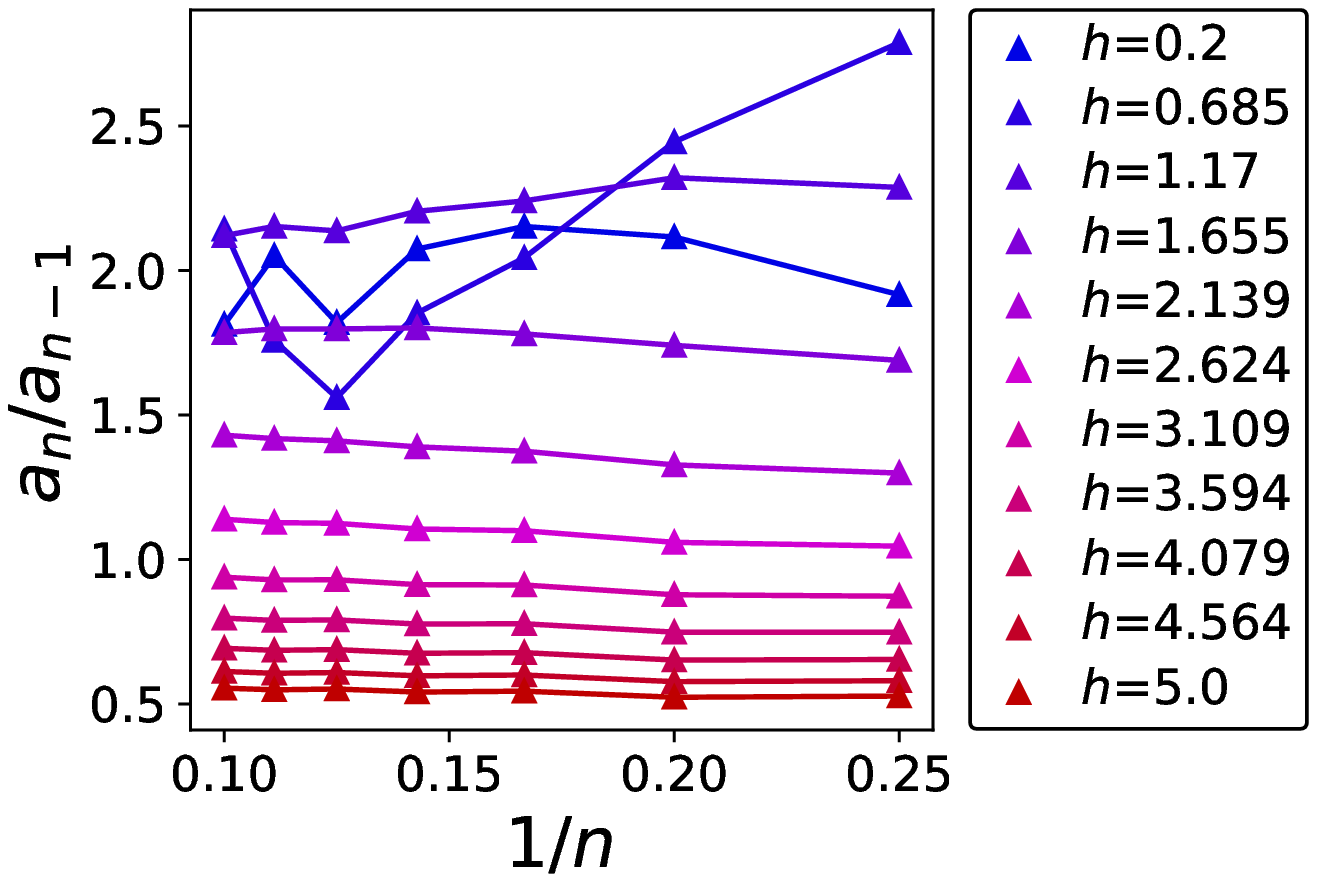}}
\caption{Plots of ratios $a_n/a_{n-1}$ as a function of $1/n$ for a representative range of $h$-values with $n$ ranging from 2 to 10.  For large values of $h$, these plots are all relatively flat.  For values of $h\lesssim h_c$, the intercept of a linear regression of the values shown here will yield a value of $1/p_c$ that is $<1$.  This implies an absence of any phase transition in the physically relevant range of $p$-values in the system.  
Subsequently, the $h$ at which the computed $p_c$ becomes unity gives an approximation of the pure system critical point $h_c$.  Additionally, for sufficiently small values of $h$, the plots become very nonlinear and irregular, indicating that the convergence of NLCE breaks down below some $h_M$.  
This is to be expected as at the multi-critical point $h_M$ critical behavior switches from being governed by the value of $h$ to being controlled by the geometry of the lattice, and the singularity is no longer a power-law for $h$ below $h_M$.}
\label{f:ratios}
\end{center}
\end{figure}

\begin{figure}[htb]
\begin{center}
\scalebox{.55}{\includegraphics{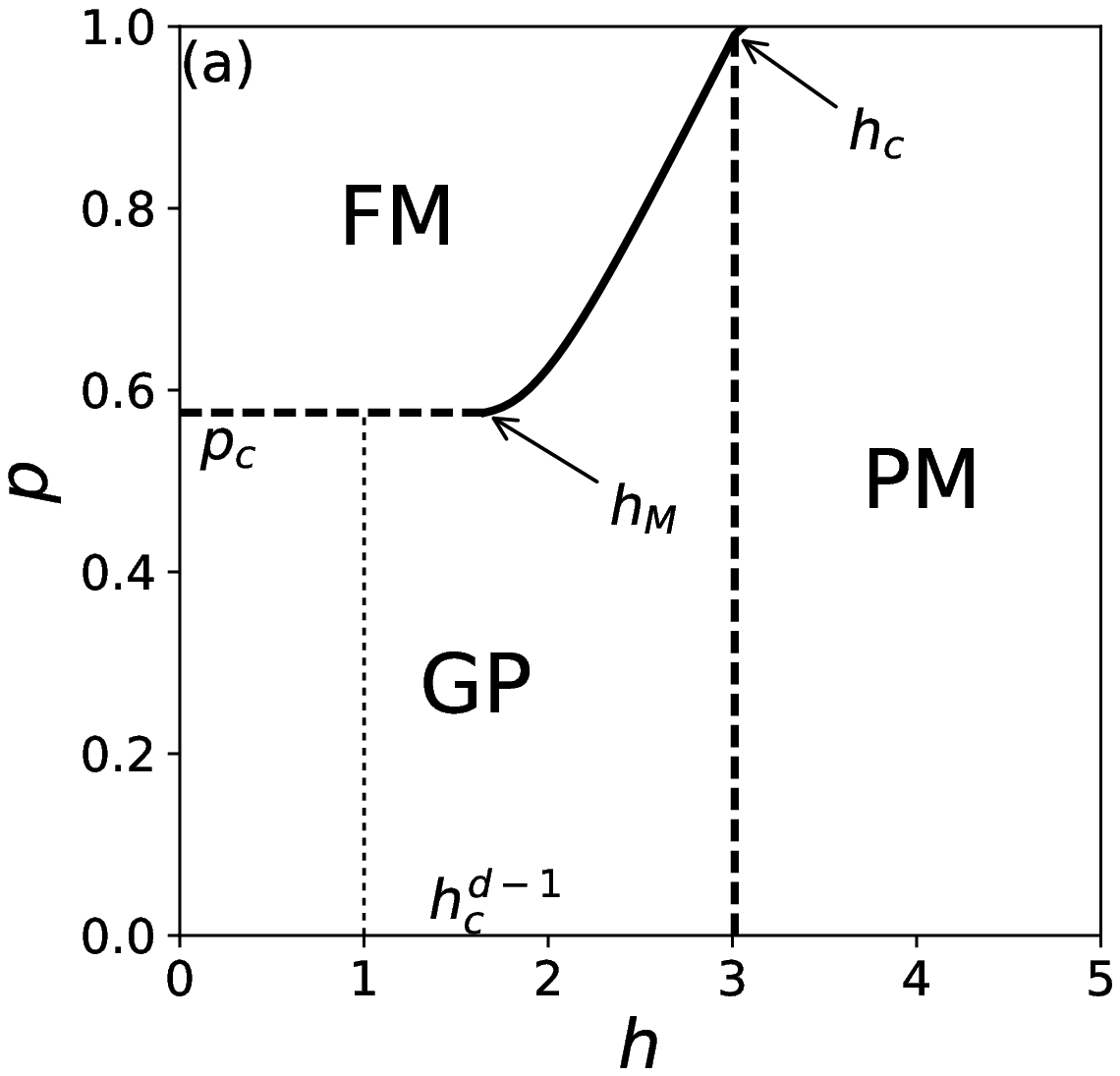}}
\scalebox{.55}{\includegraphics{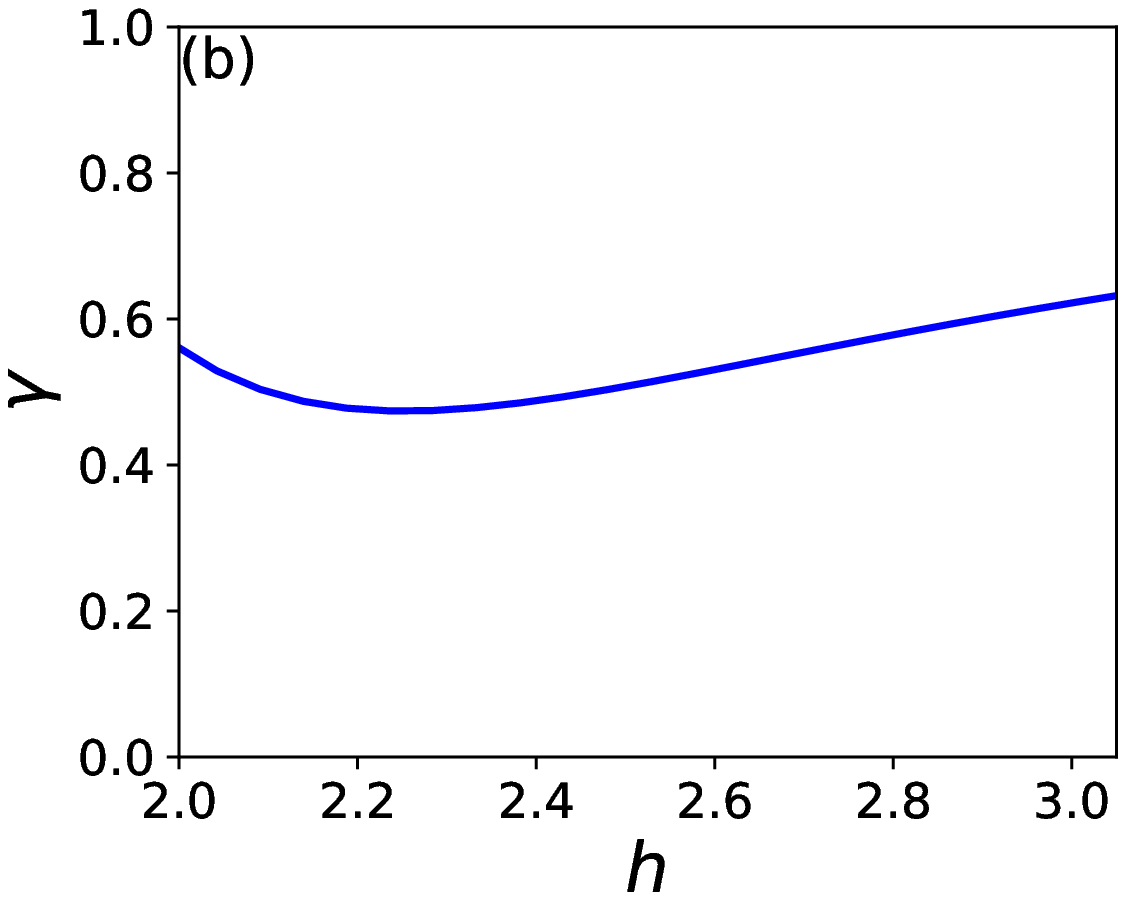}}
\caption{(a) Phase diagram, showing distinct ferromagnetic (FM), paramagnetic (PM), and Griffiths (GP) phases.  Using linear regression on data like that shown in fig. \ref{f:ratios} for many values of $h$, we established the boundary shown by the full line.  This line begins at the approximate $h_c$ of the pure system, which we find to be $3.03$, close to the known value of $3.04$.  The flat continuation of the phase boundary we expect after convergence of the NLCE breaks down for $h<h_M$ is shown by the horizontal dashed line.  This point begins at $h_M\simeq1.65$.  This line intersects the $p$ axis at the percolation probability $p_c$, which we find to be 0.58, also close to the known value of $0.59$.  The dotted line marked with $h^{d-1}_c$ shows the value of the one-dimensional critical point, a lower bound on $h_M$.  The vertical dashed line indicates a separation between the ordinary paramagnetic phase and the disordered Griffiths phase where we observe Griffiths-McCoy singularities to be present.
(b) shows the value of $\gamma$ computed from the slope of the same linear regression.  This gives the rough bounds $0.47\lesssim\gamma\lesssim0.61$.}
\label{f:phasediagram}
\end{center}
\end{figure}

Due to the simplification of NLCE to a power series as given by eq. (4), we are able to use the ratio method \cite{book} to extrapolate how the critical point $p_c$ varies as a function of $h$.  Specifically, for values of $p$ near $p_c$, we expect the structure factor $S$ to obey an asymptotic power law: $S\propto(p-p_c)^{-\gamma}$.  Subsequently, we expect the coefficients of the power series for $S$ from eq. (4) to obey (up to corrections of order $1/n^2$):
\begin{equation}
\frac{a_n}{a_{n-1}}=\frac{1}{p_c}\Big(1+\frac{\gamma-1}{n}\Big).
\end{equation}
A plot of these ratios is shown in fig. \ref{f:ratios} for a range of values of $h$.
One can estimate $p_c$ at a given $h$ from the intercept of a regression of these ratios computed for that $h$.  We use this method to compute $p_c$ as a function of $h$. The resulting phase diagram is shown in fig. \ref{f:phasediagram}(a).

In addition to this, the slope of the linear regression ratios can be used to approximate the critical exponent $\gamma$.  
Note that this is not the standard $\gamma$ exponent as we are not calculating the susceptibility. The susceptibility
itself develops strong Griffiths-McCoy singularities and hence cannot be used to study the critical point by a series
expansion method \cite{singh-young}.
In the region where this method converges well and is physically meaningful, $\gamma$ is approximately constant in the range ($0.5$ to $0.6$), within our limited accuracy, as shown in fig. \ref{f:phasediagram}(b).
The susceptibility at this transition is known to have an activated dynamical scaling behavior with a divergent dynamical
critical exponent $z$ \cite{senthil}, however, the equal time structure factor can still have a power-law behavior.
We compare our results with the double epsilon expansion of Boyanovsky and Cardy \cite{boyanovsky}, who considered the problem 
of impurities correlated along some directions. Our problem corresponds to the case of two physical dimension plus one 
correlated dimension for a total of three dimensions. Setting $\epsilon_d=1$ and $\epsilon=4-d=1$ into the expressions of
Boyanovsky and Cardy \cite{boyanovsky}, one obtains the divergence of the structure factor exponent to be $\gamma = (2-\eta-z)\nu \approx 0.73$.
Given the level of accuracy of our calculations and that this is leading order in the epsilon expansions, the agreement
is not unreasonable. We are not aware of previous numerical estimates for the exponents for the divergence of the structure
factor. 
%Fluctuation due to what?  elaborate?  How does this relate to any other results for this exponent, if there are any?
As the phase diagram flattens out the series loses convergence. This is because the system no longer has a power-law 
singularity. We estimate that the phase diagram is flat below $h_M\approx 1.65$. Again, we are not aware of previous accurate
estimates of $h_M$, which is not easy to obtain in Quantum Monte Carlo Simulations \cite{ikegami}.

\subsection{Griffiths-McCoy Singularities}
Griffiths-McCoy (GM) singularities occur in disordered quantum models where the pure system would be ordered.  Rare, ordered regions can locally mimic the ordered phase of the pure system.  
%Cite Griffiths
Here we focus on the low $p$ and $h$ regime---the Griffiths phase on the disordered side---as non-dilute regions in an otherwise highly dilute lattice.  This region is shown in fig. \ref{f:phasediagram}(a) labeled GP. There are Griffiths singularities also
on the ordered side, but we will not study them here as in the absence of a mean-field the NLCE will not converge there.  
%For small values of $h$, these regions can occupy a ferromagnetic phase independent of the behavior of the rest of the lattice.  

\begin{figure}[htb]
\begin{center}
\scalebox{.336}{\includegraphics{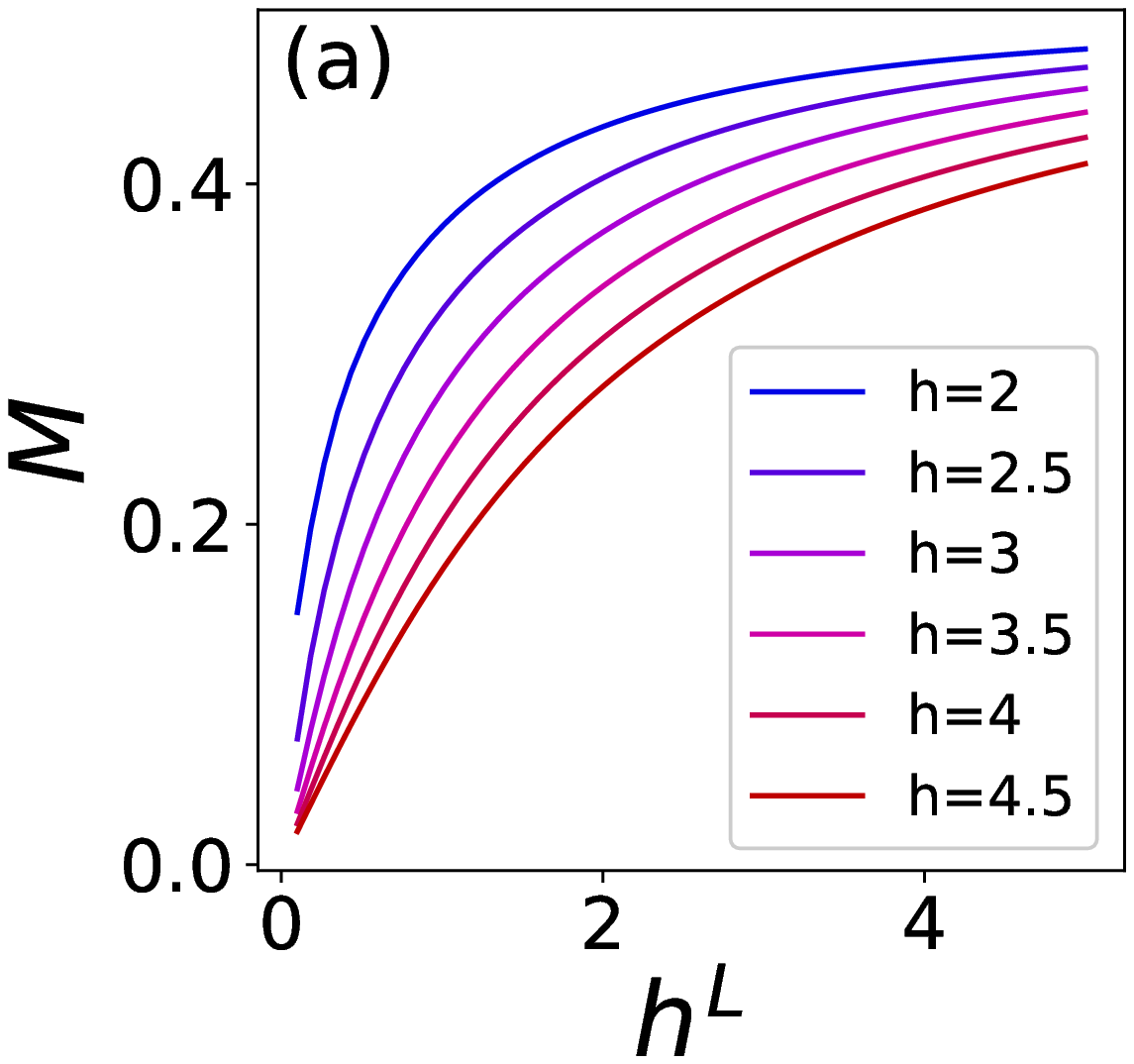}}
\scalebox{.336}{\includegraphics{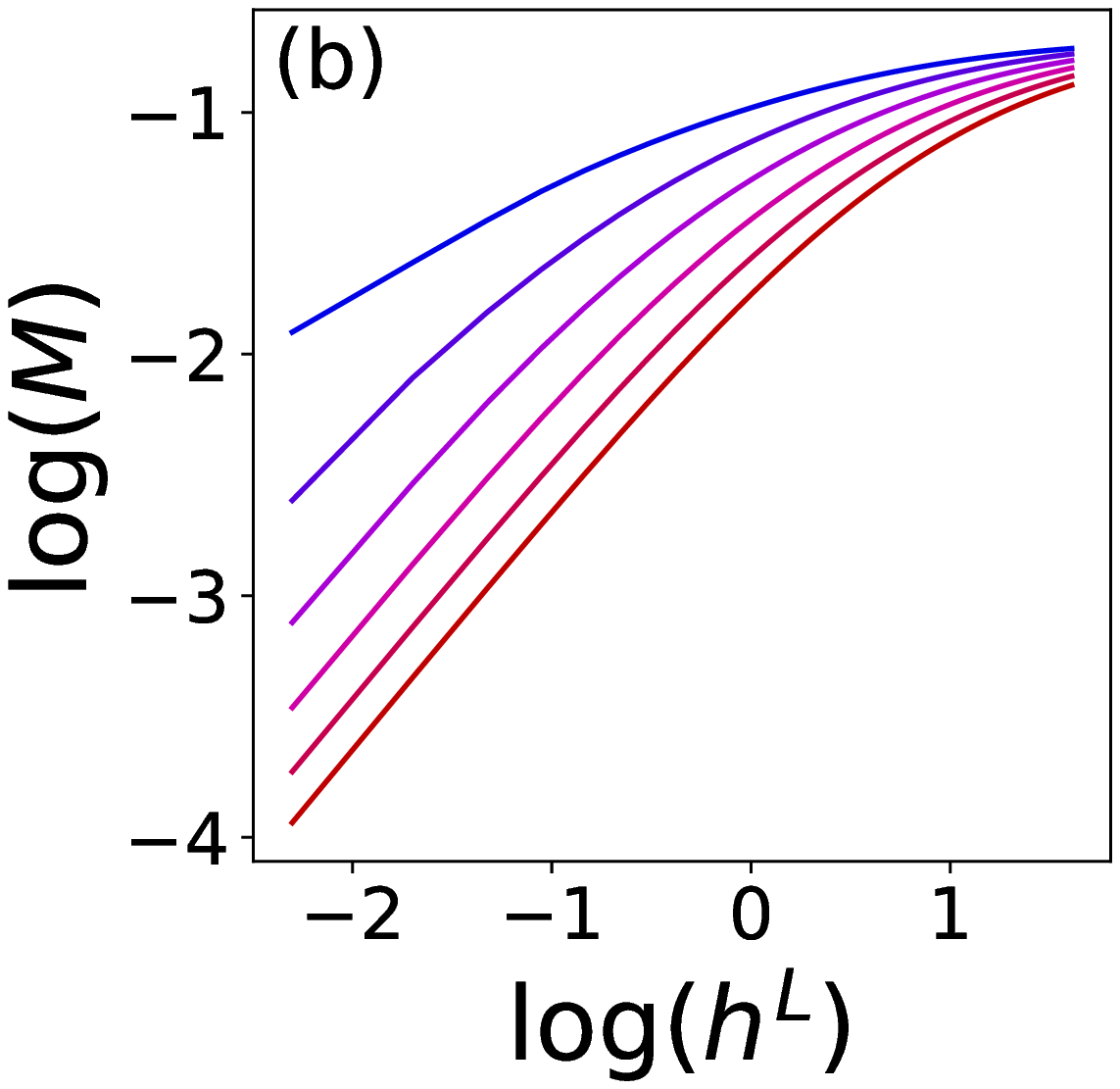}}
\caption{Both plots show computations for $p=0.5$, well below the percolation threshold $p_c=0.59$ and in the Griffiths phase for small $h$.  Computations were done to 10th order in the number of sites.  (a) shows magnetization $M$ plotted as a function of longitudinal field $h_L$ for a representative range of $h$-values.  Below the critical point $h_c=3.04$, curvature begins to develop in the small-$h_L$ limit.  This can be better visualized in (b), showing a log-log plot of the same quantities.  For $h>h_c$, the slopes of the log-log plot are about 1, while for values $h<h_c$, they are noticeably less than 1.}
\label{f:Magnetization}
\end{center}
\end{figure}

\begin{figure}[htb]
\begin{center}
\scalebox{.55}{\includegraphics{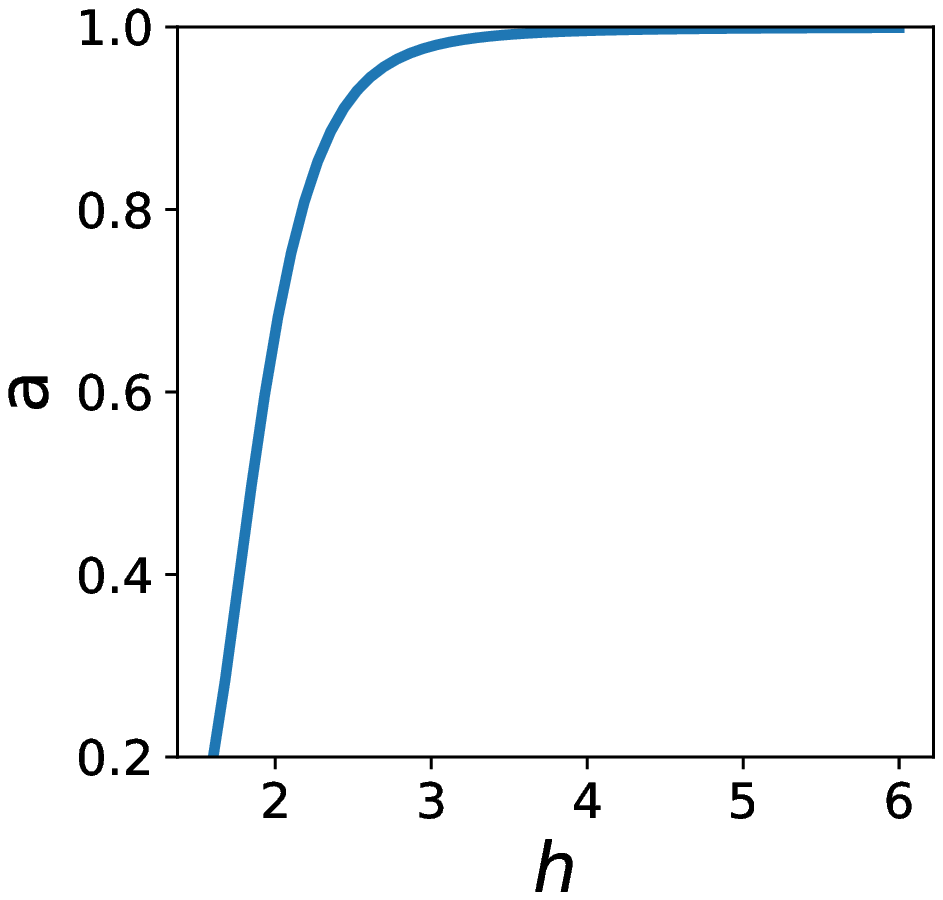}}
\caption{Plot of the exponent $a$ as a function of $h$ for $p=0.5$.  For $h>h_c$, $a\simeq1$ as is typical of the pure system.  For $h<h_c$, $a$ begins decreasing continuously with $h$, indicating the influence of GM singularities.}
\label{f:MSlope}
\end{center}
\end{figure}

We expect GM singularities to manifest in the behavior of the magnetization $M$ in a system with a longitudinal external field $h_L$:
%Why?
\begin{equation}
\mathscr{H}=J\sum\limits_{\langle i,j\rangle}\epsilon_i\epsilon_j\sigma_i^z\sigma_j^z+h\sum\limits_i\epsilon_i\sigma_i^x+h_L\sum\limits\epsilon_i\sigma_i^z.
\end{equation}
The magnetization as a function of longitudinal field of a pure ferromagnet in its paramagnetic phase is linear in the limit of small $h_L$.  However, below the percolation threshold in our dilution model, the magnetization vs $h_L$ curve develops curvature for values of $h$ below $h_c$ as shown in fig.~\ref{f:Magnetization}.
%%%%%%%%%%%%%%%%%%%%%%%%%%%%%%%%%%%%%%%%%%%%%%%%%%%%%%%%%
%NOT THE FINAL FIGURE
%%%%%%%%%%%%%%%%%%%%%%%%%%%%%%%%%%%%%%%%%%%%%%%%%%%%%%%%%
Specifically, the magnetization obeys some nonlinear power law $M\sim {h_L}^a$.  To quantify how the exponent varies with $h$, we use a linear fit of a log-log plot of $M$ and $h_L$ to compute $a$ for a range of values of $h$.  As shown in fig. \ref{f:MSlope}, for $h<h_c$, this varies continuously as a function of $h$.
%Should I include curves for p=.75 as well?

\begin{figure}[htb]
\begin{center}
\scalebox{.45}{\includegraphics{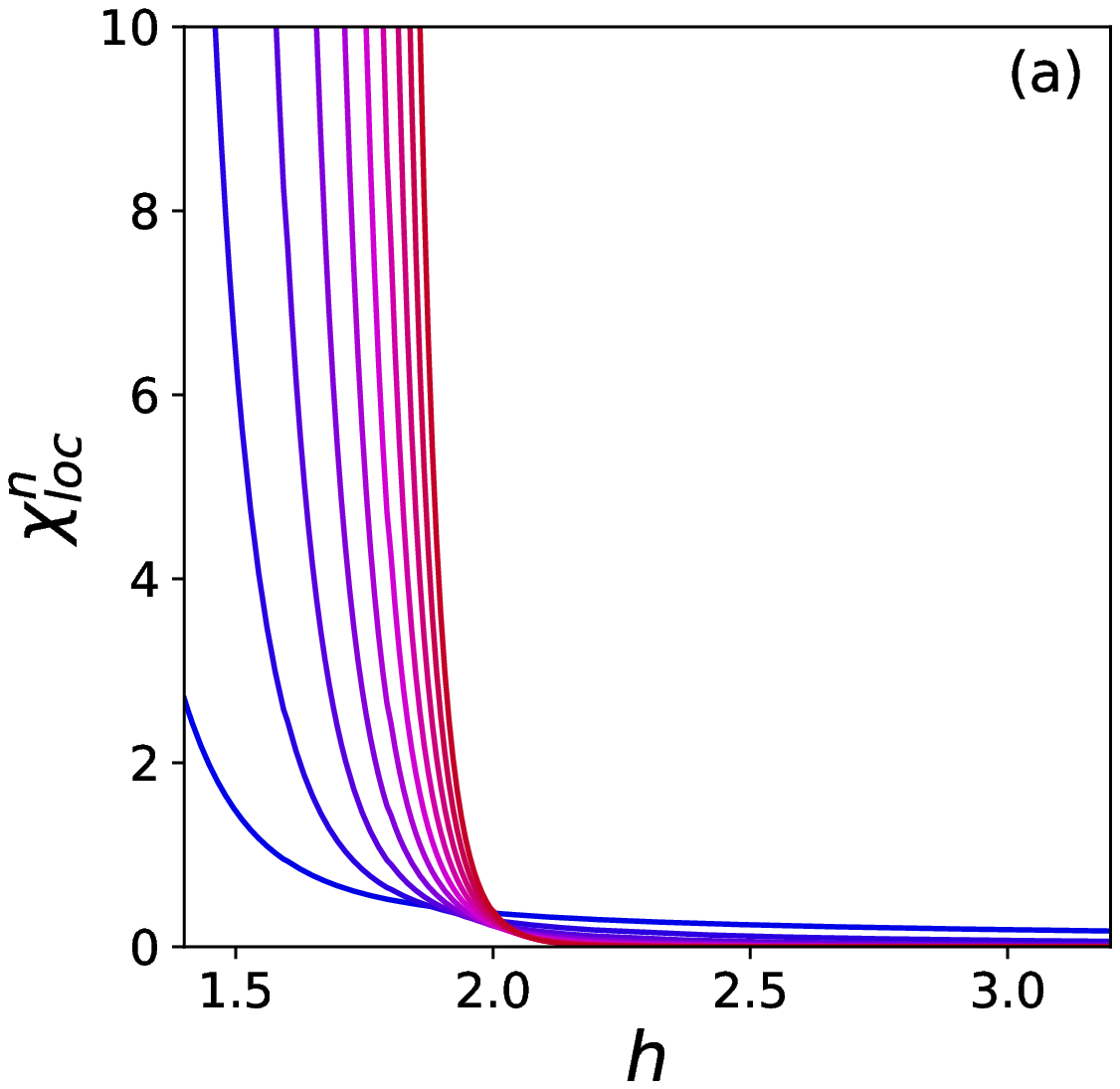}}
\scalebox{.55}{\includegraphics{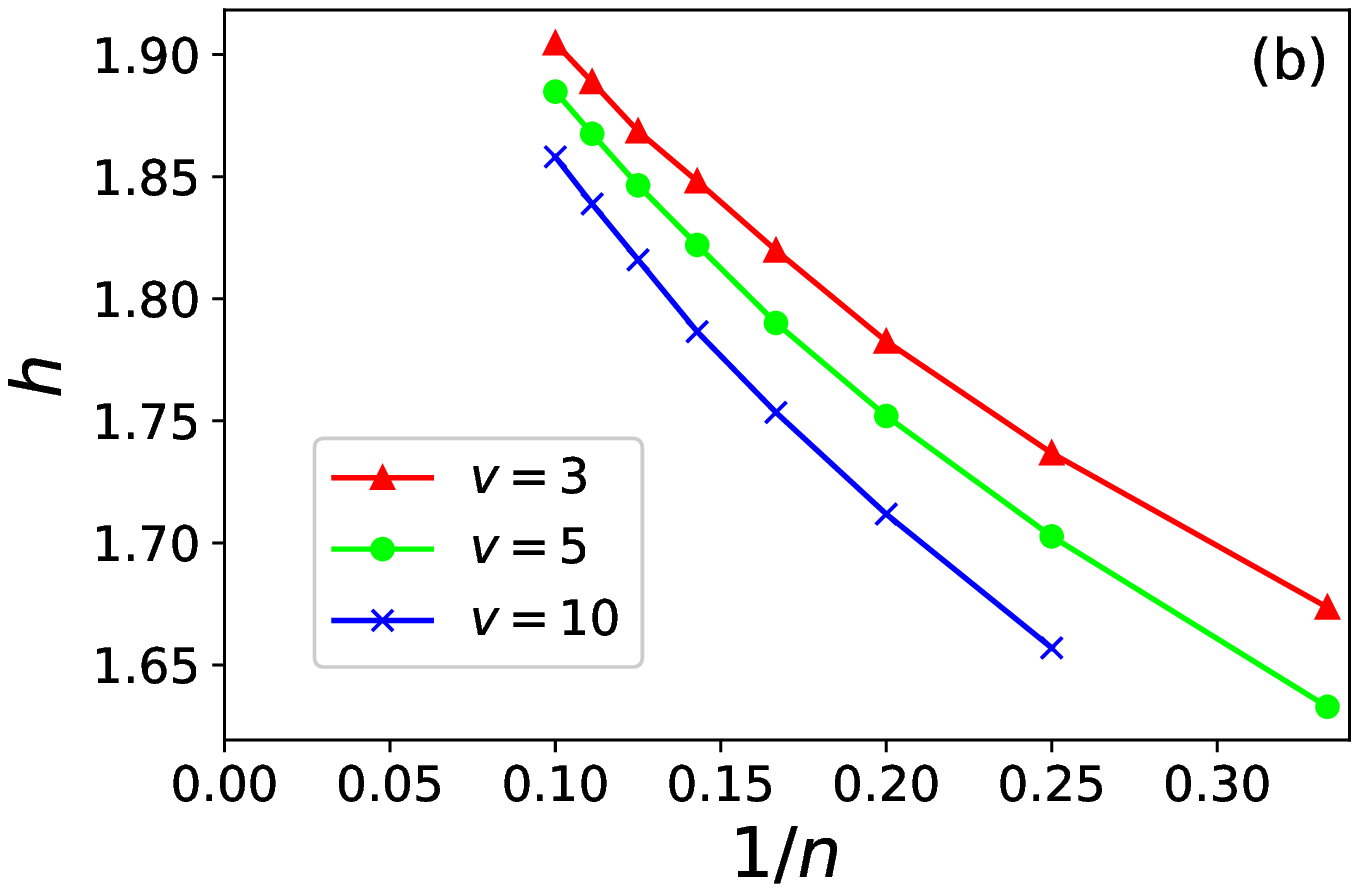}}
\caption{(a) shows plots of the moments of the local susceptibility $\chi_\text{loc}^n$ shown for $p=0.5$ for $n$ ranging from 1 to 10, with smaller $n$ values appearing in blue and larger $n$ values in red.  All values were computed to 10th order in the number of sites.
(b) shows a series of points where each moment crosses a selection of values $v$, indicating roughly where each moment begins to diverge, plotted as a function of $1/n$.  As $n$ is increased, these plots begin to curve up pushing them towards $h_c$.}
\label{f:Moments}
\end{center}
\end{figure}

As an additional indication of the influence of GM singularities, we consider the probability distribution of the local susceptibility $\chi_\text{loc}=\sum_i\chi_i$, with the one-site susceptibility defined by adding a one-site longitudinal term to the Hamiltonian:
\begin{equation}
\mathscr{H}=J\sum\limits_{\langle i,j\rangle}\epsilon_i\epsilon_j\sigma_i^z\sigma_j^z+h\sum\limits_i\epsilon_i\sigma_i^x+h_L\epsilon_i\sigma_i^z.
\end{equation}
The one-site susceptibility is then given by:
\begin{equation}
\chi_i=-\frac{\partial^2 E_0}{{\partial h_L}^2}.
\end{equation}
We study this distribution by examining its moments $\chi^n_\text{loc}\equiv\sum_i\chi_i^n$.  For small values of $n$, the value of $h$ at which the moments begin to diverge is typically much smaller than the pure system critical point $h_c$, but as $n$ is increased, the point moves closer to $h_c$.  Curves for the moments plotted over a range of values of $h$ and their points of divergence as a function of $1/n$ are shown in fig. \ref{f:Moments}(a) and (b) respectively.  This movement of the divergence point is indicative of tails in the probability distribution of $\chi_\text{loc}$ induced by the presence of GM singularities. Theoretically, the singularities set in as soon as one goes past $h_c$ of the pure system.  However, near $h_c$ they are weak and consequently only some high moments may diverge.  As one goes deeper into the Griffiths phase, the singularities become stronger and as a result, lower moments diverge as well.

\section{Conclusion}
In this work, we have used NLCE to compute the magnetization, structure factor, and susceptibility of the zero temperature dilute quantum transverse-field Ising model.  
In analyzing the pure system, we demonstrated the efficacy of the NLCE at computing the magnetization as a function of transverse field strength $h$ by adding to the Hamiltonian a mean-field term coupled to the boundary of each cluster and imposing a self-consistency constraint on the strength of the coupling and the  magnetization.  
This regulates the convergence of the NLCE in the ordered phase for the structure factor and susceptibility. It yields an approximation that converges reasonably in both the low and high $h$ regimes. Though, in the absence of more sophisticated
sequence extrapolations, it is not accurate enough to study critical phenomena. 

In the problem with dilution, NLCE turns into
a series expansion in the dilution parameter. Hence, we used a series extrapolation method to study the asymptotic behavior of the structure factor to compute $p_c$ as a function of $h$, leading to the phase boundary.  This also gave us good estimate of the pure system critical point at $h_c\simeq3.03$ and the percolation threshold $p_c\simeq0.58$, in reasonable agreement with known results of $3.04$ and $0.59$ resp.  Additionally, we found the multi-critical point at which the phase boundary flattens to be at $h_M\simeq1.65$, above the known lower bound of $1$. For the random system we found the structure factor to diverge with
an exponent $\gamma$ of $0.47\lesssim\gamma\lesssim0.61$, whose accuracy is difficult to gauge but is not far from
the Boyanovsky-Cardy values of $0.73$.

For small values of $p$ and $h$, we found numerical evidence for Griffiths-McCoy singularities in the behavior of the magnetization as a function of longitudinal field and moments of the local susceptibility.  In low $p$ regions of the paramagnetic phase, the slope $a$ of the magnetization as a function of small $h_L$ remains roughly constant at $1$ for all values of $h$. 
However, at values below $h_c$ we found that $a$ continuously diminished as a function of $h$, implying that the magnetization becomes non-linear.
This behavior mirrors observation from experiments, such as in \cite{experiments}.  
Additionally, in this region, the moments of the local susceptibility diverge at points $h<h_c$, with higher moments diverging for values of $h$ closer to $h_c$, indicating the presence of tails in the probability distribution of the local susceptibility.  Both of these effects evince a change in behavior when crossing below $h_c$ in the low-$p$ regime, demonstrating the existence of the Griffiths phase dominated by Griffiths-McCoy singularities.

\section*{Acknowledgements}
This work was supported by the NSF and U.C. Davis through an REU program by the US National Science Foundation Grant Number
PHY-1560482.

\end{document}